\documentclass[12pt]{article}
\usepackage{graphicx}

\begin{document}

\title{\textbf{Dynamical analysis on heavy-ion fusion reactions near Coulomb barrier}}

\author{Zhao-Qing Feng$^{1,2}$, Gen-Ming Jin$^{1}$, Feng-Shou Zhang$^{3}$}
\date{}
\maketitle

\begin{center}
$^{1}${\small \emph{Institute of Modern Physics, Chinese Academy of
Sciences, Lanzhou 730000, China}}  \\[0pt]
$^{2}${\small \emph{Gesellschaft f\"{u}r Schwerionenforschung mbH
(GSI), D-64291 Darmstadt, Germany}} \\[0pt]
$^{3}${\small \emph{Institute of Low Energy Nuclear Physics,
Beijing Normal University, \\[0pt]
Beijing 100875, China}}
\end{center}

\begin{abstract}
The shell correction is proposed in the improved isospin dependent
quantum molecular dynamics (ImIQMD) model, which plays an important
role in heavy-ion fusion reactions near Coulomb barrier. By using
the ImIQMD model, the static and dynamical fusion barriers,
dynamical barrier distribution in the fusion reactions are analyzed
systematically. The fusion and capture excitation functions for a
series of reaction systems are calculated and compared with
experimental data. It is found that the fusion cross sections for
neutron-rich systems increase obviously, and the strong shell
effects of two colliding nuclei result in a decrease of the fusion
cross sections at the sub-barrier energies. The lowering of the
dynamical fusion barriers favors the enhancement of the sub-barrier
fusion cross sections, which is related to the nucleon transfer and
the neck formation in the fusion reactions.
\end{abstract}


\section{Introduction}

Heavy-ion fusion dynamics at energies near and below the Coulomb
barrier has been an important subject in nuclear physics for more
than 20 years $\cite{Ba98}$, which is involved in not only exploring
several fundamental problems such as quantum tunneling etc, also
investigating nuclear physics itself, such as nuclear structure,
synthesis of superheavy nuclei etc. The experimental fusion
excitation functions can be well reproduced by the various coupled
channel methods, which include the couplings of the relative motion
to the nuclear shape deformations, vibrations, and nucleon-transfer.
However, the coupled channel models still have some difficulties in
describing the fusion cross sections for very heavy symmetric
systems, in which the neck formation will play an important role.
The microscopic dynamical description of the fusion reactions is
very necessary. Microscopic transport theories such as isospin
dependent quantum molecular dynamics (IQMD) based on QMD model
$\cite{Ai91}$ or isospin dependent Boltzmann Uehling Uhlenbeck
(IBUU) model based on BUU theory $\cite{Ue33}$ are suitable for
describing the dynamical process of the fusion reactions, which have
been used successfully to investigate heavy-ion collisions at
intermediate energies $\cite{Ch98,Zh99,Li98}$. However, these models
have some difficulties for studying the fusion reactions near
Coulomb barrier, because of some unphysical nucleon emissions in the
simulation process of projectile and target approaching. Wang et al.
had made an important improvement to construct stably initial
nucleus in the QMD model $\cite{Wa04}$, in which the interaction
potential was derived from the usual Skyrme interaction and the
parameters were adjusted by fitting experimental root-mean-square
radii, average binding energies etc. Recently, fermionic molecular
dynamics (FMD) model has also been used to investigate the fusion
reactions of oxygen isotopes $\cite{Ne07}$.

Based on IQMD model, the interaction potential, nucleon's fermionic
nature and two-body collision have been improved systematically in
ImIQMD model $\cite{Wa04,Fe05}$. In this paper, the shell correction
is further considered, and its influence on the fusion cross
sections is analyzed and compared with available experimental data.
We present a dynamical analysis on the static and dynamical fusion
barriers, the dynamical barrier distribution, and the fusion (light
and intermediate systems) and capture (heavy systems) excitation
functions and compare them with experimental data.

In Sec. $2$ we give a description on the ImIQMD model. Calculated
results of the fusion dynamics and the fusion cross sections are
given in Sec. $3$. In Sec. $4$ conclusions are discussed.

\section{Model description}

The same as the QMD or IQMD model, the wave function for each
nucleon in ImIQMD is represented by a Gaussian wave packet
\begin{equation}
\psi_{i}(\mathbf{r},t)=\frac{1}{(2\pi
L)^{3/4}}\exp\left[-\frac{(\mathbf{r}-\mathbf{r}_{i}(t))^{2}}{4L}\right]\cdot
\exp\left(\frac{i\mathbf{p}_{i}(t)\cdot\mathbf{r}}{\hbar}\right).
\end{equation}
Here the $\mathbf{r}_{i}(t)$, $\mathbf{p}_{i}(t)$ are the centers of
the $i$th nucleon in the coordinate and momentum space,
respectively. The $L$ is the square of the Gaussian wave packet
width, which depends on the size of nucleus. The total N-body wave
function is assumed as the direct product of the coherent states, in
which the anti-symmetrization is neglected. After performing Wigner
transformation for Eq. (1), we get the Wigner density as
\begin{eqnarray}
f(\mathbf{r},\mathbf{p},t)=\sum_{i}f_{i}(\mathbf{r},\mathbf{p},t),
\\
f_{i}(\mathbf{r},\mathbf{p},t)=\frac{1}{(\pi\hbar)^{3}}\exp\left[-
\frac{(\mathbf{r}-\mathbf{r}_{i}(t))^{2}}{2L}-\frac{(\mathbf{p}-\mathbf{p}_{i}(t))^{2}\cdot
2L}{\hbar^{2}}\right].
\end{eqnarray}
The density distributions in the coordinate and the momentum space
are given by
\begin{eqnarray}
\rho(\mathbf{r},t)=\int f(\mathbf{r},\mathbf{p},t) d\mathbf{p}=
\sum_{i}\frac{1}{(2\pi L)^{3/2}}\exp\left[-
\frac{(\mathbf{r}-\mathbf{r}_{i}(t))^{2}}{2L}\right], \\
g(\mathbf{p},t)=\int f(\mathbf{r},\mathbf{p},t) d\mathbf{r}=
\sum_{i}\left(\frac{2L}{\pi\hbar^{2}}\right)^{3/2}\exp\left[-
\frac{(\mathbf{p}-\mathbf{p}_{i}(t))^{2}\cdot 2L}{\hbar^{2}}\right],
\end{eqnarray}
respectively, where the sum runs over all nucleons in the systems.
Here we have considered the uncertainty relation.

The time evolutions of the nucleons in the system under the
self-consistently generated mean-field are governed by Hamiltonian
equations of motion, which are derived from the time dependent
variational principle and read $\cite{Ai91,Ha98}$
\begin{eqnarray}
\dot{\mathbf{p}}_{i}=-\frac{\partial H}{\partial\mathbf{r}_{i}},
\quad \dot{\mathbf{r}}_{i}=\frac{\partial
H}{\partial\mathbf{p}_{i}}.
\end{eqnarray}
The total Hamiltonian $H$ consists of the kinetic energy, the
effective interaction potential and the shell correction part as
follows:
\begin{equation}
H=T+U_{int}+U_{sh}.
\end{equation}
For the kinetic energy we can get it from the total wave function
\begin{equation}
T=<\Phi|\frac{\widehat{p}^{2}}{2m}|\Phi>=\int
f_{i}(\mathbf{r},\mathbf{p},t)d\mathbf{r}d\mathbf{p}=
\sum_{i}\left(\frac{\mathbf{p}^{2}_{i}(t)}{2m_{i}}+\frac{3\hbar^{2}}{8m_{i}L}\right).
\end{equation}
Here the first term is the classical kinetic energy. The second term
arises from the Gaussian width in momentum space, which is usually
omitted in QMD or IQMD calculation. It has the value 7.8 MeV if we
take $m_{i}$=938 MeV/c$^{2}$ and L=2 fm$^{2}$.

The effective interaction potential is composed of the Coulomb
interaction and the local interaction
\begin{equation}
U_{int}=U_{Coul}+U_{loc}.
\end{equation}
The Coulomb interaction potential is written as
\begin{equation}
U_{Coul}=\frac{e^{2}}{4}\sum_{i,j,j\neq
i}\frac{1}{r_{ij}}(1-t_{zi})(1-t_{zj})erf(r_{ij}/\sqrt{4L})-
\frac{3}{4}\left(\frac{3}{\pi}\right)^{1/3}e^{2}\int\rho_{p}^{4/3}d\mathbf{R},
\end{equation}
where the $t_{zi}$ is the $z$th component of the isospin degree of
freedom for the $i$th nucleon, which is equal to -1 and 1 for proton
and neutron, respectively. The
$r_{ij}=|\mathbf{r}_{i}-\mathbf{r}_{j}|$ is the relative distance of
two nucleon. The second term on the right side is the exchange term
with $\mathbf{R}=(\mathbf{r}_{i}+\mathbf{r}_{j})/2$ only for
protons, which is important for light nucleus.

In the ImIQMD model, the local interaction potential is derived
directly from the Skyrme energy-density functional
$\cite{Br85,Ch97}$, which is expressed as
\begin{equation}
U_{loc}=\int V_{loc}(\rho(\mathbf{r}))d\mathbf{r}.
\end{equation}
The local potential energy-density functional reads $\cite{Zh06}$
\begin{equation}
V_{loc}(\rho)=\frac{\alpha}{2}\frac{\rho^{2}}{\rho_{0}}+
\frac{\beta}{1+\gamma}\frac{\rho^{1+\gamma}}{\rho_{0}^{\gamma}}+
\frac{g_{sur}}{2\rho_{0}}(\nabla\rho)^{2}+\frac{g_{sur}^{iso}}{2\rho_{0}}
[\nabla(\rho_{n}-\rho_{p})]^{2}+(A\rho^{2}+B\rho^{1+\gamma}+C\rho^{8/3})\delta^{2}+
g_{\tau}\rho^{8/3}/\rho_{0}^{5/3},
\end{equation}
where the $\rho_{n}$, $\rho_{p}$ and $\rho=\rho_{n}+\rho_{p}$ are
the neutron, proton and total densities, respectively, and the
$\delta=(\rho_{n}-\rho_{p})/(\rho_{n}+\rho_{p})$ is the isospin
asymmetry. The first two terms are the bulk energy term, which are
used in the usual QMD or IQMD model. The surface term, surface
symmetry term and bulk symmetry term are from the third to the fifth
term in turn. The last term is the effective mass term, which is
related to the effective mass of nucleon together with the third
part of the fifth term. Comparing with the standard Skyrme
interaction $\cite{Br85,Ch97}$, the spin-orbit term is neglect in
the ImIQMD model, which gives the shell correction in the binding
energies. We will use a phenomenological expression to embody the
shell effect in the ImIQMD model. The coefficients in Eq. (12) are
related to the Skyrme parameters as
\begin{eqnarray}
\frac{\alpha}{2}=\frac{3}{8}t_{0}\rho_{0}, \quad
\frac{\beta}{1+\gamma}=\frac{t_{3}}{16}\rho_{0}^{\gamma}, \\
\frac{g_{sur}}{2}=\frac{1}{64}(9t_{1}-5t_{2}-4x_{2}t_{2})\rho_{0},
\\
\frac{g_{sur}^{iso}}{2}=-\frac{1}{64}[3t_{1}(2x_{1}+1)+t_{2}(2x_{2}+1)]\rho_{0},
\\
g_{\tau}=\frac{3}{80}(3t_{1}+5t_{2}+4x_{2}t_{2})\left(\frac{3}{2}\pi^{2}\right)^{2/3}\rho_{0}^{5/3}.
\end{eqnarray}
The parameters in the symmetry energy term are given by
\begin{eqnarray}
A=-\frac{1}{8}(2x_{0}+1)t_{0}, \quad B=-\frac{1}{48}(2x_{3}+1)t_{3}, \\
C=-\frac{1}{12}\left(\frac{3}{2}\pi^{2}\right)^{2/3}[t_{1}(2x_{1}+1)-t_{2}(2x_{2}+1)].
\end{eqnarray}
The $t_{0}, t_{1}, t_{2}, t_{3}$ and $x_{0}, x_{1}, x_{2}, x_{3}$
are the parameters of Skyrme force. In this work, we only consider
the first term in the bulk symmetry energy density in Eq. (12) and
take the parameter $A=a_{s}/(2\rho_{0})$, which corresponds to the
form of the linear dependence of the symmetry energy. In Table 1 we
list the ImIQMD parameters related to several typical Skyrme forces.
In the calculation we take the Skyrme parameter SLy6, which can give
the good properties from finite nucleus to neutron star
$\cite{Ch97}$.

\begin{table}
\caption{ImIQMD parameters and properties of symmetric nuclear
matter for Skyrme effective interactions} \vspace*{-10pt}
\begin{center}
\def\temptablewidth{0.8\textwidth}
{\rule{\temptablewidth}{1pt}}
\begin{tabular*}{\temptablewidth}{@{\extracolsep{\fill}}cccccccc}
&Force              &SkM*  &SIII  &SkP  &RATP   &SLy4  &SLy6 \\
\hline
&$\alpha$ (MeV)     &-317.4 &-122.7  &-356.2 &-259.2 &-298.7 &-295.7  \\
&$\beta$  (MeV)     &249.0  &55.2    &303.0  &176.9  &220.0  &216.7   \\
&$\gamma$           &7/6    &2       &7/6    &1.2    &7/6    &7/6  \\
&$g_{sur}$(MeV fm$^{2}$) &21.8 &18.3 &19.5   &25.6   &24.6   &22.9 \\
&$g_{sur}^{iso}$(MeV $fm^{2}$) &-5.5 &-4.9 &-11.3 &0.0 &-5.0 &-2.7 \\
&$g_{\tau}$ (MeV)   &5.9  &6.4  &0.0 &11.0 &9.7 &9.9 \\
&$a_{s}$ (MeV)      &30.1 &28.2 &30.9 &29.3 &32.0 &32.0 \\
&$\rho_{\infty}$ (fm$^{-3}$) &0.16 &0.145 &0.162 &0.16 &0.16 &0.16 \\
&$m_{\infty}^{\ast}/m$ &0.79 &0.76 &1.00 &0.67 &0.70 &0.69 \\
&$K_{\infty}$ (MeV)    &216  &355  &200 &239 &230 &230 \\
\end{tabular*}
{\rule{\temptablewidth}{1pt}}
\end{center}
\end{table}

The shell correction energy in the ImIQMD is written as a
Woods-Saxon form:
\begin{equation}
U_{sh}=\frac{1+x}{N}\sum_{i=1}^{N}\frac{E_{sh}}{1+\exp[(\mathbf{r}_{i}-R_{0})/a]},
\end{equation}
where the shell correction energy of ground-state nucleus $E_{sh}$
is given by the method used in Droplet model $\cite{Ro84}$. The
$\mathbf{r}_{i}$ is the radial coordinate in the center of mass
system for the $i$th nucleon. The $N, R_{0}, a$ are the total number
of nucleons, nucleus radius and dispersion width, respectively,
which has the relation $R_{0}=1.16N^{1/3}$. The coefficient $x$ is a
correction factor coming from the Woods-Saxon form, which is
determined by
\begin{equation}
x=\frac{1}{R_{0}}\int_{r_{i}}^{r_{f}}\frac{dr}{1+\exp[(r-R_{0})/a]}\approx
1.62\frac{a}{R_{0}}.
\end{equation}
We take $a=0.65$ fm in the calculation. The $r_{i}$ and $r_{f}$ are
the radial positions at $0.9\rho_{0}$ and $0.1\rho_{0}$ as shown in
Fig. 1 (a) for nucleus $^{208}$Pb with the Fermi distribution. In
Fig. 1 (b) we give a comparison of the radial density distributions
of the Fermi form, rectangular, SHF with SKM* force and the time
evolutions in the ImIQMD model. Combining Eq.(6) and Eq.(19), the
shell correction slightly constrains the nucleon motion around the
surface of a nucleus, which affects the fusion dynamics at near
Coulomb barrier energies.

\begin{figure}
\begin{center}
{\includegraphics*[width=0.5\textwidth]{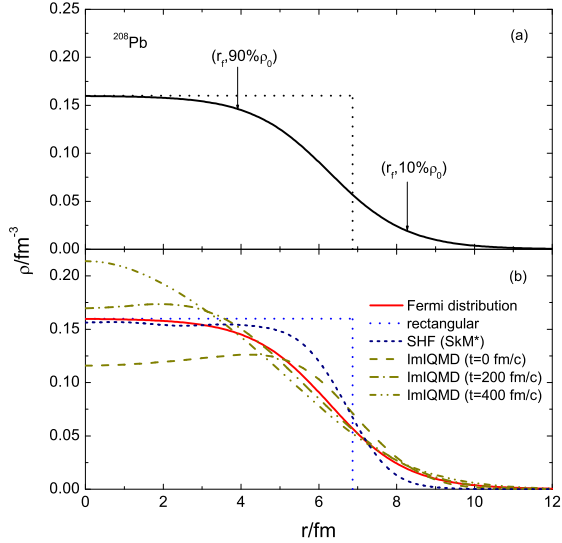}}
\end{center}
\caption{Comparisons of the radial density distributions of the
Fermi form, rectangular, Skyrme Hartree-Fock (SHF) with SkM*
parameters and the time evolutions in ImIQMD model.}
\end{figure}

In the dynamical process of two nuclei approaching, we deal with the
shell correction of the colliding system with a switch function
method $\cite{Fe05,Fe05b}$ which is used in the classical molecular
dynamics simulation $\cite{Le96}$. The shell correction energy of
the system in the dynamical evolution is written as
\begin{equation}
U_{sh}=\sum_{i=1}^{2}\frac{1+x_{i}}{N_{i}}\sum_{j=1}^{N_{i}}\frac{E_{sh}^{i}}{1+\exp[(\mathbf{r}_{j}-R_{0}^{i})/a]},
\end{equation}
where the labels $i, j$ denote the sum over the colliding nuclei and
the nucleons of the considering nucleus, respectively. The factors
$E_{sh}^{i}$ and $R_{0}^{i}$ are given by
\begin{equation}
E_{sh}^{i}(R)=E_{i}S(R)+E_{c}(1-S(R))
\end{equation}
and
\begin{equation}
R_{0}^{i}=R_{i}S(R)+R_{c}(1-S(R))
\end{equation}
respectively. Here the $E_{i,c}$ and $R_{i,c}$ are the shell
correction energies and the radii for projectile ($i=1$), target
($i=2$) and compound nucleus (c), respectively. The switch function
is expressed as
\begin{equation}
S=C_{0}+C_{1}\chi+C_{2}\chi^{2}+ C_{3}\chi^{3}+C_{4}\chi^{4}+
C_{5}\chi^{5},
\end{equation}
where $\chi=\left(\frac{R-R_{cf}}{R_{ci}-R_{cf}}\right)$ and $R$ is
the distance of the centers between projectile and target. The
quantities $R_{ci}$ and $R_{cf}$ are the distances at the initial
time and at the compound nucleus formation. In the calculation, we
set $R_{ci}=10+R_{1}+R_{2}$ fm and $R_{cf}=R_{c}-R_{2}$ fm. The
coefficients $C_{0}$, $C_{1}$, $C_{2}$, $C_{3}$, $C_{4}$ and $C_{5}$
are taken to be 0, 0, 0, 10, -15 and 6, respectively. Fig. 2 shows
the time evolutions of the shell correction energy for the double
magic nuclei $^{48}$Ca and $^{208}$Pb using Eq. (19) and in the
reaction $^{48}$Ca+$^{208}$Pb$\rightarrow$ $^{256}$No using Eq.
(21). For the single nucleus evolution in ImIQMD model, the
calculated shell correction energies approach the empirical formula
$\cite{Ro84}$ (-5.48 MeV for $^{48}$Ca and -8.93 MeV for
$^{208}$Pb). In the approaching process of two colliding nuclei, the
shell correction energy of the system evolves from the sum of
projectile and target to the one of the compound nucleus.

\begin{figure}
\begin{center}
{\includegraphics*[width=0.8\textwidth]{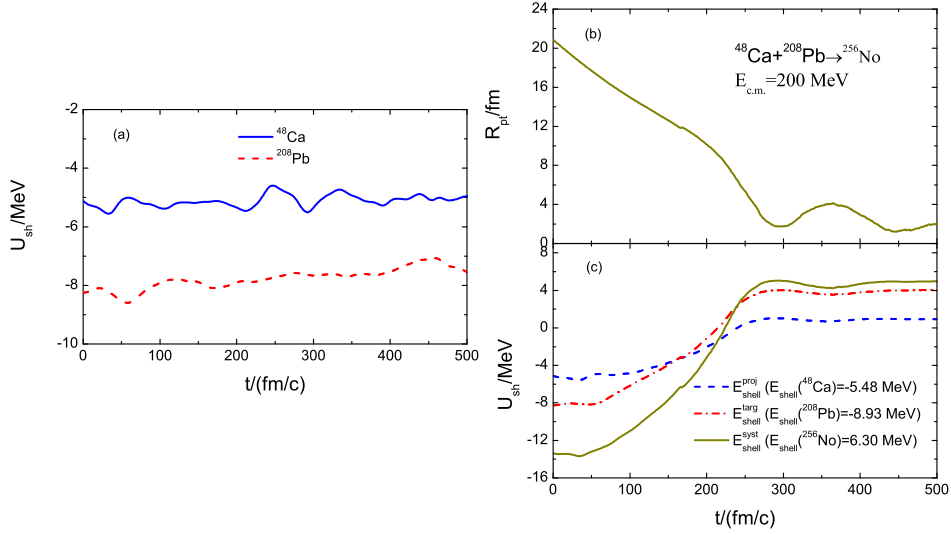}}
\end{center}
\caption{The time evolutions of the shell correction energies for
the double magic nuclei $^{48}$Ca and $^{208}$Pb (a), the distance
of centers between projectile and target (b) and the shell
correction energies of projectile, target and system in the reaction
$^{48}$Ca+$^{208}$Pb$\rightarrow$ $^{256}$No (c) at the incident
center-of-mass (c.m.) energy 200 MeV.}
\end{figure}

The phase space constraint method is also introduced in the ImIQMD
model, which is proposed by Papa et al. in order to improve the
nucleon's fermionic nature $\cite{Pa01}$. The one-body occupation
probability is given by
\begin{equation}
\overline{f}_{i}=\sum_{j}\delta_{s_{i}s_{j}}\delta_{\tau_{i}\tau_{j}}
\int_{h^{3}}f_{j}(\mathbf{r},\mathbf{p},t)d\mathbf{r}d\mathbf{p},
\end{equation}
where the quantities $s_{i}$ and $\tau_{i}$ represent the quantum
numbers of spin and isospin for the $i$th nucleon, respectively. The
integration is performed within the phase space volume $h^{3}$ at
the center position $\mathbf{r_{i}},\mathbf{p_{i}}$. In the
dynamical evolution, we timely examine the occupation probability
with the condition $\overline{f}_{i}\leq 1$. Otherwise, a series of
two-body scattering are performed to reduce the phase space
occupation probability.

The main difficulties of the usual QMD and IQMD model in the
description of heavy-ion fusion reactions near Coulomb barrier are
to construct a stable nucleus over the reaction time of the compound
nucleus formation (typically 10$^{-21}$ $\sim$ 10$^{-20}$ s). For
that Maruyama et al. proposed a cooling method in the study of the
fusion reaction $^{16}$O+$^{16}$O $\cite{Ma90}$. The stability of a
single nucleus is obviously improved in the ImIQMD model. We give an
examination of the root-mean-square radii and the average binding
energies for the magic nuclei as shown in Fig. 3. The stabilities
maintain 800 fm/c for the selected nuclei. It is necessary for
investigating the fusion dynamics of light and intermediate nuclei
or the capture process of heavy colliding systems. For the complete
fusion of heavy systems, especially the synthesis of superheavy
nuclei, or the damped reactions of two very heavy colliding nuclei
$\cite{Wa05}$, the stability of the single nucleus needs to be keep
longer time.

\begin{figure}
\begin{center}
{\includegraphics*[width=0.8\textwidth]{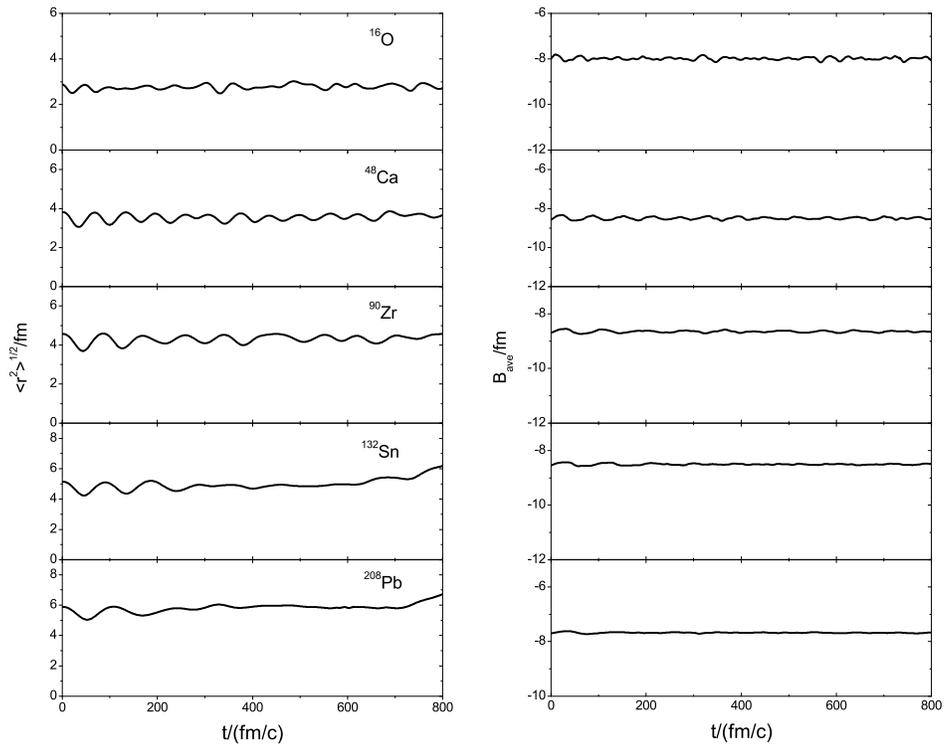}}
\end{center}
\caption{The time evolutions of the root-mean-square radii and the
average binding energies for the magic nuclei $^{16}$O, $^{48}$Ca,
$^{90}$Zr, $^{132}$Sn and $^{208}$Pb.}
\end{figure}

\section{Results and discussions}

In this section the ImIQMD model is applied to analyze the fusion
dynamics at energies near Coulomb barrier. It is well known that the
enhancement of the sub-barrier fusion cross sections is related to
the various couplings of the relative motion to the nuclear shape
vibrations, deformations and nuclear transfer degrees of freedom.
The fusion barrier at various energies and the dynamical barrier
distribution are investigated systematically using the ImIQMD model.
The influence of the shell effect on the fusion cross sections is
analyzed. The fusion and capture excitation functions for a series
of reaction systems are calculated and compared them with
experimental data.

\subsection{Fusion barrier and dynamical barrier distribution}

The interaction potential $V(R)$ of two colliding nuclei as a
function of the distance $R$ between their centers is defined as
$\cite{Br68}$
\begin{equation}
V(R)=E_{pt}(R)-E_{p}-E_{t}.
\end{equation}
Here the $E_{pt}$, $E_{p}$ and $E_{t}$ are the total energies of the
whole system, projectile and target, respectively. The total energy
is the sum of the kinetic energy, the effective potential energy and
the shell correction energy. In the calculation, the Thomas-Fermi
approximation is adopted for evaluating the kinetic energy
$\cite{Fe05}$. The static and dynamical interaction potentials are
calculated according to Eq. (26) as shown in Fig. 4 for head on
collisions of the reaction systems $^{48}$Ca+$^{238}$U and
$^{36}$S+$^{90,96}$Zr. The static interaction potential means that
the density distribution of projectile and target is always assumed
to be the same as that at initial time, which is a diabatic process
and depends on the collision orientations and the mass asymmetry of
the reaction systems. The total density of the system is taken as
the sum of the individual nucleus in the overlapping region. For a
comparison, the proximity results $\cite{My00}$ and the adiabatic
barrier as mentioned in Ref. $\cite{Si01}$ are also shown in the
figure (left panel) and the corresponding barrier heights are
indicated for the various cases. However, for a realistic heavy-ion
collision, the density distribution of the whole system will evolve
with the reaction time, which significantly depends on the incident
energy and impact parameter for a given system. In the calculation
of the dynamical potentials, we only pay attention to the fusion
events as shown later, which give the fusion dynamical barrier. At
the same time, stochastic rotation is performed for different
simulation events. One can see that the heights of the dynamical
barriers are reduced gradually with decreasing the incident energy
and increasing the neutron number of the target. The lowering of the
dynamical fusion barrier is in favor of the enhancement of the
sub-barrier fusion cross sections, which can give a little of
information that the cold fusion reactions are also suitable to
produce superheavy nuclei. We can understand the microscopic process
of the lowering from the neck formation which derives from the
nucleon transfer and the dynamical deformation in the fusion
reactions $\cite{Fe05c}$.

\begin{figure}
\begin{center}
{\includegraphics*[width=0.8\textwidth]{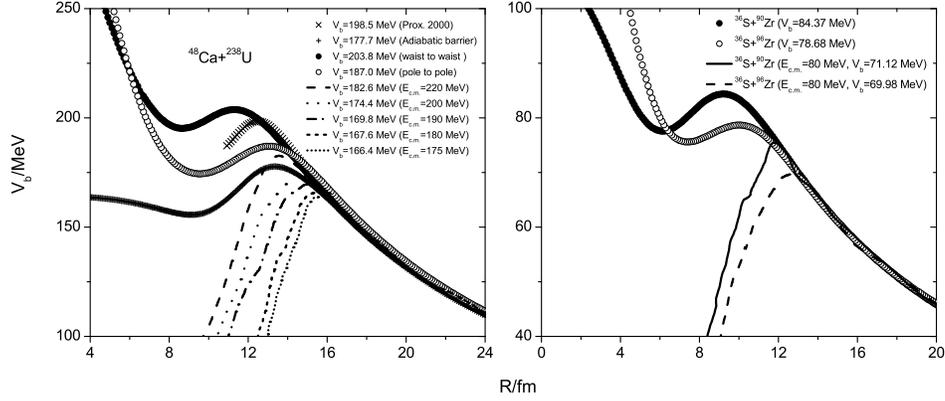}}
\end{center}
\caption{Comparison of the static fusion barriers (including
pole-to-pole and waist-to-waist collisions), adiabatic barrier,
proximity results and dynamical barriers at different incident c.m.
energies for the reaction $^{48}$Ca+$^{238}$U (left panel), as well
as system dependence of the static and dynamical fusion barriers
(right panel).}
\end{figure}

The dynamical fusion barrier is calculated by averaging the fusion
events at a given energy and an impact parameter. To explore more
information on the fusion dynamics, we also investigate the
dynamical barrier distribution as shown in Fig. 5 for head on
collisions of the reaction $^{36}$S+$^{90}$Zr at incident energies
80 MeV and 85 MeV (static barrier $V_{b}=84.37$ MeV), respectively.
The distribution moves towards the lower barrier region with
decreasing the incident energy. A number of fusion events are
located at the sub-barrier region, which is favorable to sub-barrier
fusion reactions. Rowley et al. proposed a method of extracting the
fusion barrier distribution from the second energy derivative of the
fusion excitation function $\cite{Ro91}$. A portion of the
distribution is located at the sub-barrier region from the analysis
of experimental fusion excitation functions, which may help us
understanding the sub-barrier fusion and the kind of various
couplings.

\begin{figure}
\begin{center}
{\includegraphics*[width=0.8\textwidth]{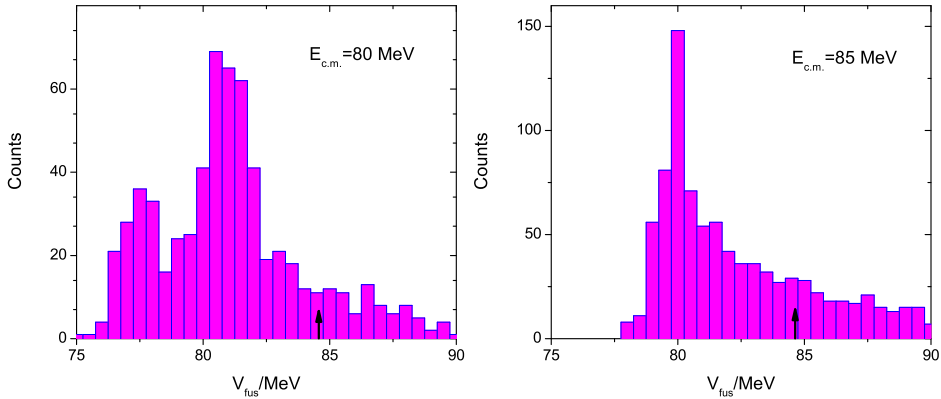}}
\end{center}
\caption{Dynamical barrier distribution for the reaction
$^{36}$S+$^{90}$Zr at incident energies 80 MeV and 85 MeV,
respectively. Arrows show the static fusion barriers.}
\end{figure}

\subsection{Fusion and capture excitation functions}

After constructing the stable events of projectile and target, the
simulation of the fusion reaction can be performed. Stochastic
rotation around their centers of mass for each nucleus by a Euler
angle is made for every event at a given incident energy and an
impact parameter. The simulation events are set to be 200 for each
incident energy $E$ and impact parameter $b$, and be 300 at the
sub-barrier energies. The fusion cross section is calculated by the
formula
\begin{equation}
\sigma_{fus}(E)=2\pi\int_{0}^{b_{max}}bp_{fus}(E,b)db=2\pi\sum_{b=\Delta
b}^{b_{max}}bp_{fus}(E,b)\Delta b,
\end{equation}
where $p_{fus}(E,b)$ stands for the fusion probability and is given
by the ratio of the fusion events $N_{fus}$ to the total events
$N_{tot}$. The reliability of the calculated $p_{fus}(E,b)$ is
estimated by the value $\cite{Ma90}$
\begin{equation}
\Delta P_{f}=
1.64\left[\frac{N_{fus}(N_{tot}-N_{fus})}{N_{fus}^{3}}\right]^{1/2}.
\end{equation}
In the calculation the step of the impact parameter is set to be
$\Delta b=0.5$ fm. We also use the same procedure in the definition
of the fusion event in Ref. $\cite{Wa04}$ that the coalesced
one-body density can survive through one or more rotations of the
composite system or through several oscillations of its radius. The
capture cross section of heavy colliding system is also calculated
using Eq. (27), in which the capture event is defined as that the
distance between the centers of colliding nuclei is smaller than the
minimum position of the pocket of the static interaction potential.
In Fig. 6 we show the dependence of the fusion probability as a
function of the impact parameter on the incident energies in the
reaction $^{36}$S+$^{90}$Zr and on the system size at the energy
$E_{c.m.}$=80 MeV. One can see that the higher incident energy and
the neutron-rich system have the larger fusion probability, and the
fusion probability is reduced with increasing the impact parameter
or the relative angular momentum at each incident energy. There is
no signal on the appearance of the so-called "fusion window" as
predicted by the time dependent Hartree-Fock method $\cite{Wo82}$.

\begin{figure}
\begin{center}
{\includegraphics*[width=0.8\textwidth]{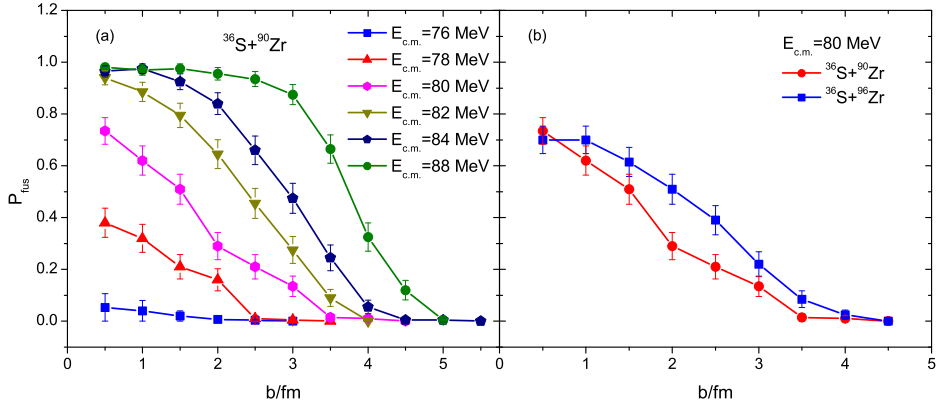}}
\end{center}
\caption{The fusion probability as a function of the impact
parameter and its dependence on the incident energies and on the
system size.}
\end{figure}

Recent experimental data showed that there was no enhancement of the
fusion cross sections in the reactions $^{48}$Ca+$^{90}$Zr or
$^{96}$Zr comparing those in the reactions $^{40}$Ca+$^{90}$Zr or
$^{96}$Zr although using more neutron-rich projectile $^{48}$Ca
$\cite{St06}$. Possible reason is explained from the fact that the
double magic nucleus $^{48}$Ca has more rigid (stronger shell
effect) structure than $^{40}$Ca. We analyzed the influence of the
shell effect on the fusion cross sections in the reaction
$^{48}$Ca+$^{90}$Zr as shown in Fig. 7. With the same initial
conditions such as the stable event and the distance of the centers
between two nuclei etc, the calculated fusion cross sections are
obviously reduced after considering the shell correction, especially
in the sub-barrier domain. The main reason is that the shell
correction potential in Eq. (7) constrains the motion of the surface
nucleons of the colliding nuclei. The experimental data can be
reproduced rather well by considering the shell correction.

\begin{figure}
\begin{center}
{\includegraphics*[width=0.5\textwidth]{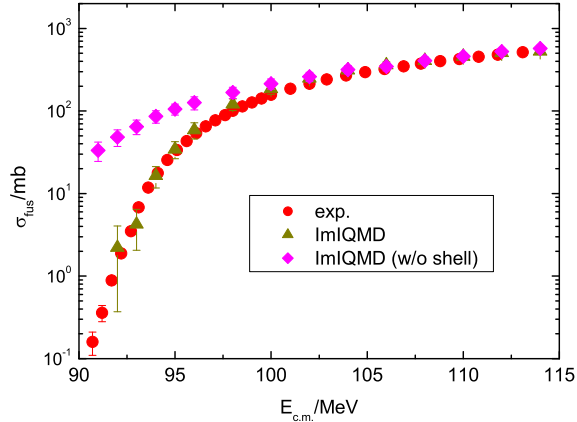}}
\end{center}
\caption{The calculated fusion cross sections with and without
considering shell effect in the reaction $^{48}$Ca+$^{90}$Zr and
compared with experimental results $\cite{St06}$.}
\end{figure}

Fig. 8 shows a comparison of the calculated fusion excitation
functions by the ImIQMD model after considering the shell
correction, one-dimensional Hill-Wheeler formula $\cite{Hi53}$ and
the experimental results for the reactions $^{46}$Ti+$^{46}$Ti
$\cite{St02}$ and $^{40}$Ca+$^{112}$Sn $\cite{Sc00}$. The
Hill-Wheeler formula underestimates the fusion cross sections at the
sub-barrier energies for the two reaction systems. There is no other
adjustable parameters in the ImIQMD model, which is a purely
dynamical process. The agreement of the calculated fusion excitation
functions with the experimental data is remarkably well within
statistical error bars. For systematically examining the reliability
of the model and exploring the fusion dynamics, we calculated the
fusion cross sections of a series of the reaction systems and
compared them with available experimental data
$\cite{St06,Bo92,St00,Mo94,We91}$ as shown in Fig. 9. One can see
that the neutron-rich combinations have the larger fusion cross
section, especially at the sub-barrier regions. As we discussed in
the section 3.1, the neutron-rich systems give the lower dynamical
fusion barriers at the same incident c.m. energy, which are
favorable to increase the fusion cross sections, especially at the
sub-barrier energies. The reactions $^{18}$O+$^{58}$Ni and
$^{16}$O+$^{60}$Ni lead to the same compound nucleus formation.
However, the system $^{18}$O+$^{58}$Ni has the larger fusion cross
sections at the sub-barrier energies. Zagrebaev explained the
enhancement owing to the positive $Q$ value of the neutron transfer
by using a simplified model $\cite{Za03}$. Therefore, the colliding
system has a little surplus energy to pass over the interaction
barrier. In the ImIQMD model, since the magic nucleus $^{16}$O has
the stronger shell effect than $^{18}$O, the transfer of the surface
nucleon is slightly constrained in the course of projectile and
target approaching, which also results in the decrease of the
sub-barrier fusion cross sections.

\begin{figure}
\begin{center}
{\includegraphics*[width=0.8\textwidth]{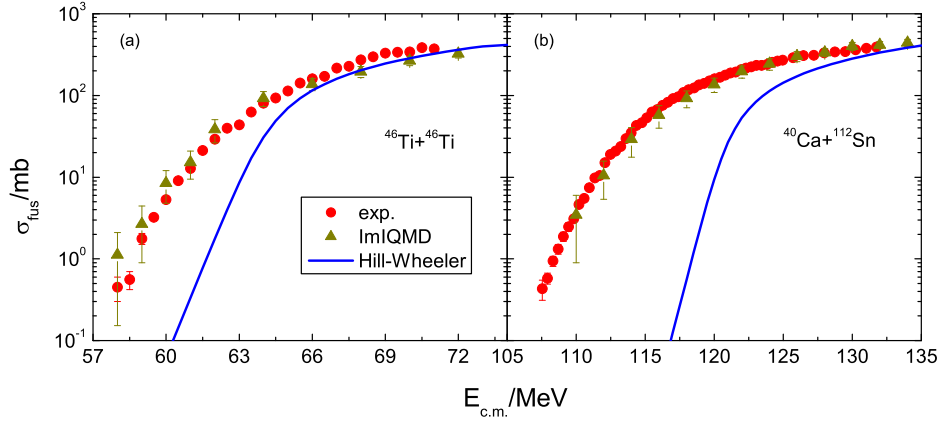}}
\end{center}
\caption{Comparisons of the calculated fusion excitation functions,
Hill-Wheeler formula $\cite{Hi53}$ and experimental data
$\cite{St02,Sc00}$ for the reactions $^{46}$Ti+$^{46}$Ti and
$^{40}$Ca+$^{112}$Sn.}
\end{figure}

\begin{figure}
\begin{center}
{\includegraphics*[width=0.8\textwidth]{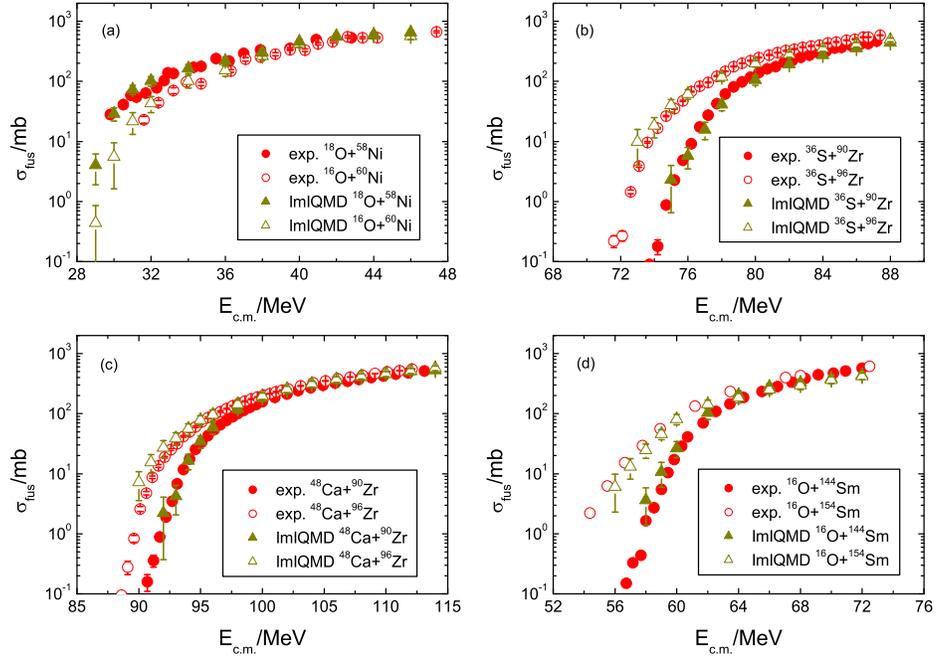}}
\end{center}
\caption{Comparisons of the calculated fusion excitation functions
and experimental data for a series of reaction systems.}
\end{figure}

The synthesis of heavy or superheavy nuclei is mainly reached by the
complete fusion reactions of two heavy colliding nuclei in
experimentally $\cite{Ho00,Og07}$. In accordance with the evolution
of the colliding system, the whole process of the compound nucleus
formation and decay is usually divided into three reaction stages,
namely the capture process of the colliding system to overcome
Coulomb barrier, the formation of the compound nucleus to pass over
the inner fusion barrier, and the de-excitation of the excited
compound nucleus against fission $\cite{Fe06}$. As the first stage
of synthesizing superheavy nuclei, the accurate calculation of the
capture cross sections and the analysis of the capture dynamics are
very important to estimate the evaporation residue cross sections
and also affect the competition of the complete fusion and the
quasi-fission. Within the framework of the ImIQMD model, we analyzed
the time evolutions of the relative angular momentum and the
distance between the centers of the projectile and target in the
$^{48}$Ca+$^{238}$U reaction at incident energy $E_{c.m.}$=200 MeV
as shown in Fig. 10. The relative angular momentum is defined as
$J=(X_{c.m.}^{p}-X_{c.m.}^{t})\sqrt{2\mu E_{c.m.}}/\hbar$ with the
reduced mass of the colliding system $\mu$ and the center of mass
distance in the $X$ directions for projectile and target which
defines the impact parameter at the initial time. The colliding
system evolves from the individual nuclei to the dinuclear system
formation, which is a process of the dissipation of the relative
angular momentum and the kinetic energy of the relative motion. In
Fig. 11 we present a comparison of the calculated capture cross
sections and the experimental data for a series of reaction systems.
The experimental data are taken from Refs.
$\cite{Da04,Ni04,Pr03,Sh87}$. One can see that the calculated
results are in good agreement with the experimental data. Fig. 12
also shows the calculated capture excitation functions for the
reactions $^{48}$Ca+$^{244}$Pu and $^{48}$Ca+$^{248}$Cm and compared
them with available experimental data $\cite{It04}$, which have been
used to synthesize superheavy elements Z=114 and Z=116 in Dubna
$\cite{Og00}$. Here, we have considered the quadrupole deformation
for the deformed nucleus at the initial sampling. The capture
process of the considered systems takes place in the evolution range
t=600-800 fm/c. Then, the composite system evolves into either
separating two fragments with a larger probability (quasi-fission)
or maintaining a whole system with a smaller probability which leads
to the compound nucleus formation. It needs a longer evolution time
and the larger simulation events to investigate the quasi-fission
and the complete fusion in the two heavy colliding nuclei. Further
works are in progress.

\begin{figure}
\begin{center}
{\includegraphics*[width=0.8\textwidth]{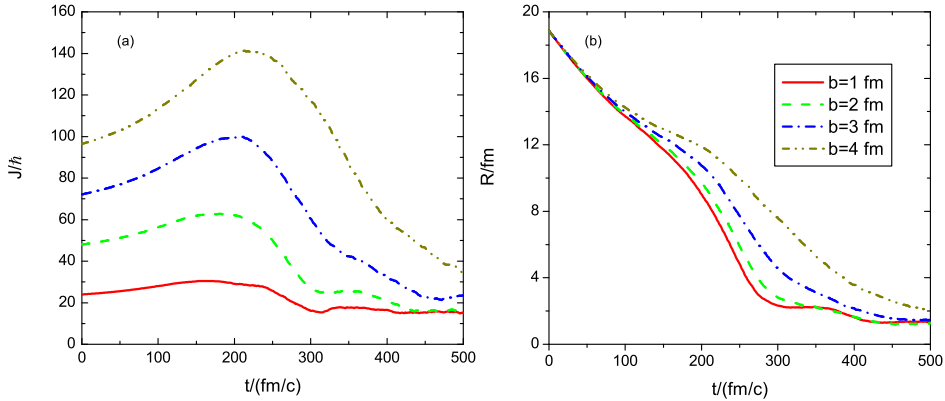}}
\end{center}
\caption{The time evolutions of the relative angular momentum and
the distance between the centers of two colliding nuclei in the
reaction $^{48}$Ca+$^{238}$U at incident energy 200 MeV.}
\end{figure}

\begin{figure}
\begin{center}
{\includegraphics*[width=0.8\textwidth]{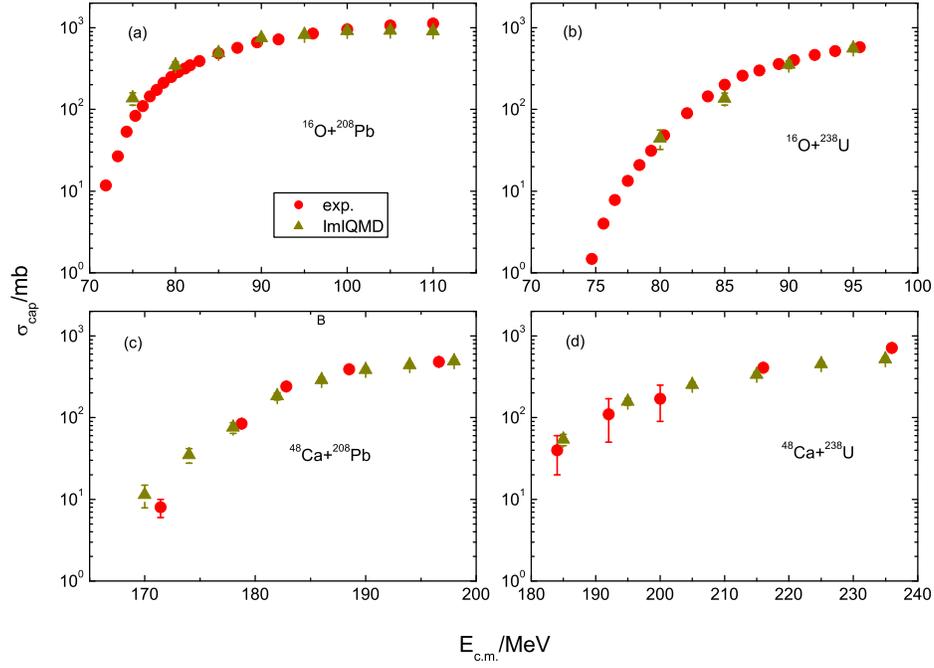}}
\end{center}
\caption{Comparisons of the calculated capture excitation functions
and experimental data for the selected reaction systems.}
\end{figure}

\begin{figure}
\begin{center}
{\includegraphics*[width=0.8\textwidth]{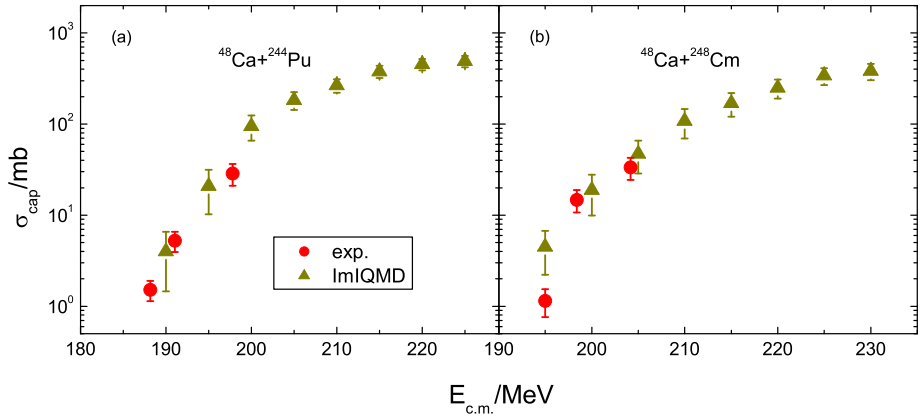}}
\end{center}
\caption{The calculated capture excitation functions for the
reactions $^{48}$Ca+$^{244}$Pu and $^{48}$Ca+$^{248}$Cm and compared
them with available experimental data $\cite{It04}$.}
\end{figure}

\section{Conclusions}

The shell correction has been further considered in the ImIQMD
model. Its influence on the fusion excitation functions is analyzed
and compared with the experimental data. By using the ImIQMD model,
the fusion dynamics in heavy-ion collisions near Coulomb barrier is
investigated systematically, such as the fusion barrier, dynamical
barrier distribution etc. The calculated excitation functions of the
fusions and captures are in good agreement with the available
experimental data. The lowering of the dynamical fusion barrier
favors the enhancement of the sub-barrier fusion cross sections. The
strong shell effect constrains the fusion of two colliding nuclei at
near barrier energies.

The physical nature of the heavy-ion fusion reactions near Coulomb
barrier is very complicated, which is a dynamical process involving
the excitation of the colliding system. Semi-classical dynamical
model can explore its dynamical behavior in a certain extent. There
are also some interesting works applying the ImIQMD model to
investigate the halo nucleus induced reactions, the fusion dynamics
(capture, quasi-fission and complete fusion) of the synthesis of the
superheavy nuclei in the massive fusion reactions, and searching
other possible ways to produce superheavy nuclei etc. For those
cases, the structure properties and the longer stability of the
initial nucleus have to be made.

\section{Acknowledgements}

One of us (Z.Q. Feng) is grateful to Profs. Zhuxia Li, Junqing Li,
Werner Scheid, Hans Feldmeier and Dr. Ning Wang for fruitful
discussions and help, and also thanks the hospitality during his
stay in GSI. This work was supported by the National Natural Science
Foundation of China under Grant No. 10475100, the Major state basic
research development program under Grant No. 2007CB815000, and the
Helmholtz-DAAD in Germany.

\end{document}